\documentclass[12pt]{article}
\usepackage{a4wide}
\usepackage{amssymb}
\usepackage{graphicx}
\begin{document}

{\renewcommand{\thefootnote}{\fnsymbol{footnote}}
\hfill  IGPG--06/7--4\\
\medskip
\hfill gr--qc/0607130\\
\medskip
\begin{center}
{\LARGE  Quantum Geometry and its\\ Implications for Black Holes\footnote{Plenary talk at ``Einstein's Legacy in the New Millenium,'' Puri, India, December 2005}}\\
\vspace{1.5em}
Martin Bojowald\footnote{e-mail address: {\tt bojowald@gravity.psu.edu}}
\\
\vspace{0.5em}
Institute for Gravitational Physics and Geometry,\\
The Pennsylvania State
University,\\
104 Davey Lab, University Park, PA 16802, USA\\
\vspace{1.5em}
\end{center}
}

\setcounter{footnote}{0}

\begin{abstract}
 General relativity successfully describes space-times at
 scales that we can observe and probe today, but it cannot be complete
 as a consequence of singularity theorems.  For a long time there have
 been indications that quantum gravity will provide a more complete,
 non-singular extension which, however, was difficult to verify in the
 absence of a quantum theory of gravity.  By now there are several
 candidates which show essential hints as to what a quantum theory of
 gravity may look like. In particular, loop quantum gravity is a
 non-perturbative formulation which is background independent, two
 properties which are essential close to a classical singularity with
 strong fields and a degenerate metric.  In cosmological and black
 hole settings one can indeed see explicitly how classical
 singularities are removed by quantum geometry: there is a
 well-defined evolution all the way down to, and across, the smallest
 scales. As for black holes, their horizon dynamics can be studied
 showing characteristic modifications to the classical behavior.
 Conceptual and physical issues can also be addressed in this context,
 providing lessons for quantum gravity in general. Here, we conclude
 with some comments on the uniqueness issue often linked to quantum
 gravity in some form or another.
\end{abstract}

\newcommand{\lP}{\ell_{\rm P}}
\newcommand{\md}{{\mathrm d}}

\section{General relativity}

General relativity successfully describes the gravitational field, in
terms of space-time geometry, on large scales. However on small scales
it is hardly being probed by direct observations. Nevertheless, there
are indirect indications, and all of them point to its failure:
singularity theorems imply not only infinite densities but even a
complete breakdown of evolution after a finite amount of proper time
for most realistic solutions \cite{SingTheo}. The best known examples
are cosmological situations and black holes.

Traditionally, such a breakdown of evolution has been interpreted as
corresponding to the beginning, or end, of the universe. In a
situation like this it is instructive to remember what is probably
best expressed by the quote ``{\em The limits of my language
mean the limits of my world.}'' \cite{LW}.
Currently, our best language to speak about the universe is general
relativity, but it clearly has its limitations. The description it
provides of the world is incomplete when curvature becomes large, and
such limitations should not be mistaken for the limits of the actual
world it is supposed to describe.

Such strong curvature regimes are reached when relevant length scales
become small such as in black holes or the early universe. One can
understand the failure of general relativity as a consequence of
extrapolating the well-known long-distance behavior of gravity to
unprobed small scales. According to general relativity, the
gravitational force is always attractive such that, at some point,
there will be nothing to prevent the total collapse of a region in
space-time to a black hole, or of the whole universe.

Moreover, this is a purely classical formulation, and further expected
modifications arise on small scales when they approach the Planck
length $\ell_{\rm P}=\sqrt{8\pi G\hbar}\approx 10^{-35}{\rm m}$.
Possible spatial discreteness, also an expectation from well-known
quantum properties, would imply a radically different underlying
geometry. These lessons from quantum mechanics can be made more
precise, although they remain at a heuristic level until they are
confirmed by a theory of quantum gravity (see, e.g.,
\cite{WS:MB}).

First, the hydrogen atom is known to be unstable classically since the
electron falls into the nucleus in a finite amount of time, a
situation not unlike that of the collapse of a universe into a
singularity. Upon quantization, however, the atom acquires a finite
ground state energy and thus becomes stable. The value,
$E_0=-\frac{1}{2} me^4/\hbar^2
\stackrel{\hbar\to0}{\longrightarrow}-\infty$, also shows the
importance of quantization and the necessity of keeping the Planck
constant non-zero. In quantum gravity one can expect on similar
dimensional grounds that densities are bounded by an inverse power of
the Planck length such as
$\lP^{-3}\:\stackrel{\hbar\to0}{\longrightarrow}\infty$ which again
diverges in the classical limit but is finite in quantum gravity.

The second indication comes from black body radiation where the
classical Rayleigh--Jeans law, which would imply diverging total
energy, is modified by quantum effects on small (wavelength)
scales. The resulting formula for its spectral energy density, due to
Planck, gives a finite total energy.

This observation gives us an indication as to what could happen when
one successfully combines general relativity with quantum theory.
According to Einstein's equations, matter energy density back-reacts
on geometry. If matter energy behaves differently on small scales due
to quantum effects, the classical attractive nature of gravity may
change to repulsion. In addition, also the non-linear gravitational
interaction itself, without considering matter, can change.

All those expectations, some of them long-standing, have to be
verified in explicit calculations which requires a {\em
non-perturbative} (due to strong fields) and {\em background
independent} (due to degenerate geometry) framework of quantum
gravity.  Non-perturbativity can be implemented by adopting a
canonical quantization, which when done in Ashtekar
variables \cite{AshVar,AshVarReell} also allows one to deal with
background independence: there are natural smearings of the basic
fields as holonomies and fluxes which do not require the introduction
of a background metric and nonetheless result in a well-defined
kinematical algebra \cite{LoopRep,ALMMT}.

\section{Loop Quantum Gravity}

Instead of using the spatial metric and extrinsic curvature for the
canonical formulation following Arnowitt, Deser and Misner \cite{ADM},
the Ashtekar formulation \cite{AshVar,AshVarReell} expresses general
relativity as a constrained gauge theory with connection
$A_a^i=\Gamma_a^i+\gamma K_a^i$ and its momenta given by a densitized
triad $E^a_i$ expressing spatial geometry. In the connection,
$\Gamma_a^i$ is the spin connection compatible with the densitized
triad and thus a function of $E_a^i$, while $K_a^i$ is extrinsic
curvature. The Barbero--Immirzi parameter \cite{AshVarReell,Immirzi}
$\gamma>0$ is free to choose and does not have classical implications.

\subsection{Representation}

These basic fields $(A_a^i,E_j^b)$ 
must now be smeared in order to obtain a well-defined Poisson algebra
(without delta functions) suitable for quantization. However, the
common 3-dimensional smearing of all basic fields is not possible
because the spatial metric is now dynamical and there is no other
background metric. Fortunately, Ashtekar variables allow a natural
smearing without background geometry by using
\begin{eqnarray*}
 \mbox{\em holonomies} && h_e(A)=P\exp\int_e\tau_i
A_a^i\dot{e}^a{\rm d}t \quad \mbox{for edges }e\\
 \mbox{and {\em fluxes}} && F_S(E)=\int_S \tau^i  E^a_in_a{\rm
d}^2y \quad \mbox{for surfaces }S
\end{eqnarray*}
as smeared objects. Indeed, their Poisson algebra closes with
well-defined structure constants. Edges $e$ and surfaces $S$ arising
in this definition play the role of labels of the basic objects and
chosen freely in a given 3-manifold $\Sigma$. Alternatively,
edges and surfaces can be introduced as abstract sets with a relation
showing their intersection behavior which determines the structure
constants of the algebra.

This algebra can now be represented on a Hilbert space to define a
quantum theory. Requiring diffeomorphism covariance of the
representation, i.e.\ the existence of a unitary action of the
diffeomorphism group, even fixes essentially the available
representation to that used in loop quantum gravity
\cite{LOST,WeylRep}. It is most easily constructed in terms of states
being functionals of the connection such that holonomies become basic
multiplication operators. Starting with the simplest possible state
which does not depend on the connection at all, holonomies ``create''
spin network states \cite{RS:Spinnet} upon action. Most generally,
states are then of the form $T_{g,j,C}(A)=\prod_{v\in g} C_v\cdot
\prod_{e\in g} \rho_{j_e}(h_e(A))$ where $g$ is an oriented graph
formed by the edges $e$ used in multiplication operators. Each edge
carries a label $j$ corresponding to an irreducible SU(2)
representation and arising from the fact that the same edge can be
used several times in holonomies.  Finally, $C_v$ are contraction
matrices to multiply the matrix elements of $\rho_{j_e}(h_e(A))$ for
edges containing the vertex $v$ to a complex number. These contraction
matrices can be chosen such that the state becomes invariant under
local SU(2) gauge transformations.

\subsection{Discrete geometry}

Flux operators can be derived using the fact that they are
conjugate to holonomies, and thus become derivative
  operators on states. Replacing triad components by functional
derivatives and using the chain rule, we obtain
\begin{eqnarray*}
 \hat{F}_S f_g &=& -i\gamma\ell_{\rm P}^2 \int_S {\rm d}^2y\tau^i n_a
\frac{\delta}{\delta A_a^i(y)}
f_g(h(A))\\
 &=&-i\gamma\ell_{\rm P}^2
\sum_{e\in g}\int_S {\rm d}^2y\tau^i n_a
\frac{\delta h_e}{\delta
    A_a^i(y)}\frac{\md f_g(h)}{\md h_e}
\end{eqnarray*}
with non-zero contributions only if $S$ intersects the edges of
$g$. Moreover, each such contribution is given by the action of an
SU(2) derivative operator with discrete spectrum. The whole spectrum
of flux operators is then discrete implying, since fluxes encode
spatial geometry, discrete spatial quantum geometry. This translates
to discrete spectra also of more familiar spatial geometric
expressions such as area or volume \cite{AreaVol,Area,Vol2}.

\subsection{Dynamics}

It does, however, not directly imply discrete space-{\em time}
geometry since this requires dynamical information encoded, in a
canonical formulation, in the Hamiltonian constraint. There are
classes of well-defined operators for this
constraint \cite{RS:Ham,QSDI}, which usually change the graph of the
state they act on. Their action is therefore quite complicated in full
generality, not unexpectedly so for an object encoding the quantized
dynamical behavior of general relativity.

As usually, symmetries can be used to obtain simpler expressions while
still allowing access to the most interesting phenomena in gravity
such as cosmology or black holes. In loop quantum gravity with its
well-developed mathematical techniques \cite{ALMMT}, moreover, such
symmetries can be imposed {\em at the quantum level} \cite{SymmRed} by
inducing a reduced quantum representation from the unique one in the
full theory. This is particularly useful because in symmetric
situations usually no uniqueness theorems hold, for instance in
homogeneous models where the widely used Wheeler--DeWitt
representation is inequivalent to the representation arising from loop
quantum gravity. The loop representation in such models is then
distinguished by its relation to the unique representation of the full
theory. On the loop representation one can then construct more
complicated operators such as the Hamiltonian constraint in analogy to
the full construction. Often, the operators simplify considerably and
can sometimes be used in explicit calculations.

\subsection{Loop quantum cosmology}
\label{lqc}

Following this procedure in cosmological situations of homogeneous
spatial slices, the usual Wheeler--DeWitt equation \cite{DeWitt} is
replaced by a {\em non-singular} difference
equation \cite{Sing,IsoCosmo,HomCosmo,Spin}. For this equation, the
wave function of a universe model is uniquely defined once initial
conditions are imposed at large volume. In particular, the difference
equation continues to determine the wave function even at and beyond
places where classical singularities would occur and also the
Wheeler--DeWitt equation would stop. For a mathematical discussion of
properties of the resulting difference equations, see \cite{Encyc}.

As in the full setting, there are currently different versions (such
as symmetric and non-symmetric orderings, or other ambiguity
parameters) resulting in non-singular behavior. Some versions,
however, are singular which means that ambiguities are already
restricted by ruling them out. Nevertheless, some aspects can change
also between different non-singular versions, such as the issue of
{\em dynamical initial conditions} \cite{DynIn,Essay} which are
consistency conditions for wave functions provided by the dynamical
law rather than being imposed by hand. They arise with stronger
restrictions on wave functions in a non-symmetric ordering compared to
a symmetric one. Such ambiguities in simple models have to be
constrained by studying more complicated situations (see, e.g.,
\cite{GenFunc,GenFuncBI,PreClassBI}).  Through this
procedure, the theory becomes testable because it is highly
non-trivial that one and the same mechanism (including, e.g., the same
ordering choice) applies to all situations.

Intuitively, the behavior can be interpreted as giving a well-defined
evolution to a new branch {\em preceding the big bang}; see
\cite{WS:MB} for a general discussion. This new
branch is provided by a new, binary degree of freedom given by the
orientation of the triad. We are naturally led to this freedom since
triads occur in the background independent formulation. It is
precisely the orientation change in triad components which presents us
with two sides to classical singularities. Unlike the Wheeler--DeWitt
equation which abruptly cuts off the wave function at vanishing metric
components, loop quantum cosmology has an equation connecting the wave
function on both sides of a classical singularity. This is a very
general statement and applies to {\em all} possible initial conditions
at large volume, compatible with potential dynamical initial
conditions. It is thus independent of complicated issues such as which
of the solutions to the difference equation are normalizable in a
physical inner product (see \cite{DegFull} for more details).

For more precise information on the structure of classical
singularities in quantum gravity, however, one needs additional
constructions. For instance, the evolution of semiclassical states can
be studied in detail in some special models such as a flat isotropic
model coupled to a free, massless scalar \cite{QuantumBigBang,Param}.
Here, the classical singularity turns out to be replaced by a bounce
at small scales, connecting two semiclassical phases at larger
volume. It remains open, however, how general this scenario is when
potentials or anisotropies are included.

When a physical inner product or precise semiclassical states are
unavailable, one can make use of effective equations of the classical
type which are ordinary differential equations in coordinate time but
incorporate some prominent quantum effects. They have been introduced
in \cite{Inflation} and applied in the context of bounces in
\cite{BounceClosed,Time,KasnerBounce}. Recently, it has
been shown, using a geometrical formulation of quantum
mechanics \cite{Schilling,Josh}, that this is part of a general scheme
which agrees with effective action techniques common from quantum
field theory where both approaches can be applied \cite{EffAc,Karpacz}.
Thus, these equations allow an effective analysis of quantum theories
in the usual sense. Also those equations show bounces, sometimes even
in semiclassical regimes, but not generically. Effects can depend on
quantization choices as well as on which quantum corrections are
included. There are different types of corrections which in general
are all mixed without any one being clearly dominant. A full study,
including all possible quantum correction terms, is complicated and
still to be done \cite{Karpacz}.

All these bounce scenarios can be seen intuitively as confirmation of
our expectation that quantum gravity should contribute a repulsive
component to the gravitational force on small scales. Such repulsion
can stop the collapse of a universe and turn it into a bounce, after
which the weakening repulsion will contribute to accelerated expansion
in an inflationary scenario \cite{Inflation}.

\section{Black holes}

Unlike cosmological models, black holes require inhomogeneous
situations. There are currently several techniques to get hints for
the resulting behavior and for typical quantum effects in the physics
of black holes \cite{BlackHoles}.

\subsection{The Kantowski--Sachs model}

Inside the horizon, the Schwarzschild solution is homogeneous because
the Killing vector field which is timelike in the static outside
region turns spacelike. With the three rotational Killing vectors this
combines to a 4-dimensional symmetry group corresponding to
Kantowski--Sachs models.  Densitized triads with this symmetry can be
written as (see \cite{BHInt,ModestoConn} for details on this part)
\[
E = p_c\tau_3\sin\vartheta\frac{\partial}{\partial
x}+p_b\tau_2\sin\vartheta\frac{\partial}{\partial\vartheta}
-p_b\tau_1\frac{\partial}{\partial\varphi}
\]
such that $\det E=p_c p_b^2$ and orientation, which is important for
the singularity structure, is given by ${\rm sgn} p_c$. The sign of
$p_b$ is not relevant as there is a residual gauge transformation
$p_b\mapsto -p_b$. From such a densitized triad, the spatial metric
\[
 \md s^2 = \frac{p_b^2}{|p_c|}\md x^2+ |p_c|\md\Omega^2
\]
results. Comparison with the interior Schwarzschild metric suggests
the following identification between space-time and minisuperspace
locations: the {\em Schwarzschild singularity} at $p_c=0$
and the {\em horizon} at $p_b=0$.

When quantized, the densitized triad components become operators
\begin{equation}
 \hat{p}_b|\mu,\nu\rangle = \frac{1}{2}\gamma\lP^2\mu
 |\mu,\nu\rangle \quad,\quad \hat{p}_c|\mu,\nu\rangle=
\gamma\lP^2\nu|\mu,\nu\rangle
\end{equation}
acting on orthonormal states $|\mu,\nu\rangle$ with
$\mu,\nu\in{\mathbb R}$, $\mu\geq0$. This Hilbert space is the analog
of the spin network representation in the full theory, although most
labels disappeared thanks to the high symmetry. Also analogously to the
full theory one can construct the Hamiltonian constraint operator
which gives rise to the dynamical law
\begin{eqnarray}
&&2(\sqrt{|\nu+2|}+\sqrt{|\nu|})
\left(\psi_{\mu+2,\nu+2}- \psi_{\mu-2,\nu+2}\right)\nonumber\\
&& +(\sqrt{|\nu+1|}-\sqrt{|\nu-1|})
\left((\mu+2)\psi_{\mu+4,\nu}-
(1+2\gamma^2)\mu\psi_{\mu,\nu}+
(\mu-2)\psi_{\mu-4,\nu}\right)\nonumber\\
&&+2(\sqrt{|\nu-2|}+\sqrt{|\nu|})
\left(\psi_{\mu-2,\nu-2}-
\psi_{\mu+2,\nu-2}\right)=0
\end{eqnarray}
as a difference equation for the wave function depending on the triad
components. This is singularity free as in the isotropic case,
extending the wave function beyond the classical singularity. The
situation is more complicated, however, because the classical
minisuperspace now has two boundaries, one corresponding to the
horizon at $p_b=0$ and one at the singularity corresponding to
$p_c=0$. But only one direction can be extended given only one sign
factor from orientation. Thus, the system provides an interesting
consistency check of the general scheme by determining which boundary,
the singularity or the horizon, is removed upon quantization. The
horizon boundary should not be removed because at this place our
minisuperspace approximation breaks down. Indeed it is just the
classically singular boundary which is removed by including the sign
of $p_c$ in the analysis, providing a non-trivial test of the
singularity removal mechanism of loop quantum cosmology. This rests
crucially on the use of densitized triad variables which we are led to
naturally in a full background independent formulation. While models
also allow quantizations in terms of other variables, e.g.\ using
co-triads or metrics \cite{Modesto}, classical singularities appear at
different places of minisuperspace and general schemes of singularity
removal do then not exist.

\subsection{Evaporation}

By extrapolating the extension of the interior Schwarzschild geometry
through the classical singularity to dynamical situations in the
presence of matter one can arrive at a new paradigm for {\em black
hole evaporation} \cite{BHPara} founded on loop quantum gravity.

\begin{figure}
\begin{center}
\mbox{}\hspace{1cm}\includegraphics[height=7cm]{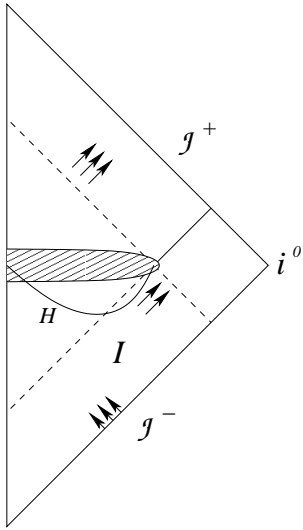}\hspace{1cm}
\parbox[b]{9cm}{\caption{Evaporating non-singular black hole. The hatched quantum
region replaces the classical singularity such that infalling matter
can propagate through it and re-emerge later. The horizon $H$
evaporates due to Hawking radiation. \label{NonSingH}}}
\end{center}
\end{figure}

As illustrated in Fig.~\ref{NonSingH}, the quantum region around the
classical black hole singularity is not a future boundary of
space-time. Correlations between infalling matter components are then
not destroyed during evaporation. Instead, matter will be able to
leave the black hole region, defined by the presence of trapped
surfaces and a horizon $H$ \cite{trapping,DynHorLett,DynHor,HorRev},
after the evaporation process restoring all correlations. There are
thus trapped surfaces which as in the usual singularity theorems of
general relativity implies geodesic incompleteness. But here, this
does not lead to a singularity; rather the space-time continuum is
replaced by a discrete, quantum geometrical structure. Only classical
concepts break down but not the quantum gravitational
description. There is then also no need to continue the horizon $H$
beyond the point where it meets the strong quantum region, although
there can well be past trapped surfaces in the future of the quantum
regime. In this sense, the picture is similar to closed horizons
enclosing a bounded space-time region which have been suggested
earlier \cite{ClosedHorHighDer,ClosedHor,ClosedHorTrapping}.

There are several underlying assumptions for this scenario which are
now being tested. For instance, for this picture space-time has to
become semiclassical again after evolving through the quantum regime
where discrete geometry is essential. Otherwise, there will be no
asymptotically flat future space-time to detect remaining
correlations. In the Schwarzschild interior discussed above, one can
verify that solutions to the difference equation have to be symmetric
under $\mu\mapsto-\mu$ as a consequence of consistency conditions in
the recurrence scheme \cite{GenFuncKS}. Note that this is not a
condition imposed on solutions but follows from the dynamical law,
similarly to dynamical initial conditions for
cosmology \cite{DynIn,Essay}. In the non-interacting, empty
Schwarzschild solution the future is thus the time reverse of the past
and in particular becomes semiclassical also to the future. Whether
semiclassical behavior in the future is also achieved in more
complicated collapse scenarios is not known so far.

\subsection{Outside the horizon}

The behavior is more complicated if matter is present or when
inhomogeneities are considered. Then, back-reaction of Hawking
radiation \cite{PredictBreakdown} on geometry and scattering of matter
leads to a future behavior different from the past of the quantum
region. For such cases, it has not been shown that space-time becomes
semiclassical after all fields are settled down.

To complete the picture, access to the outside of the horizon is needed
as well as a handle on field degrees of freedom of matter or
gravity itself. Moreover, the horizon dynamics must be understood
taking into account quantum effects. There are two main ways
to approach this complex issue:
\begin{itemize}
\item {\em Effective equations} \cite{Inflation,Perturb,Josh,EffAc}, 
which have been successful in
understanding qualitative aspects of homogeneous models (e.g.,
\cite{Inflation,BounceClosed,BounceQualitative,Oscill,InflOsc,EmergentLoop,LoopFluid,NonChaos,ChaosLQC}),
are not yet available for inhomogeneous situations, but the
homogeneous forms can be exploited in matchings.
\item {\em Midisuperspace
    models} and their quantum dynamics close to classical
    singularities or at horizons, related to properties of difference
    equations, are being developed and have already given initial
    promising insights.
\end{itemize}

\subsubsection{Matching}

Gravitational collapse is often modeled by matching a homogeneous
matter distribution (such as a star) to an inhomogeneous exterior
geometry, following the work of Oppenheimer and Snyder \cite{OppSny}.
Modeling the collapsing body by an isotropic interior solution
\[
 \md s^2=-\md t^2+\frac{a(t)^2}{(1+\frac{1}{4}r^2)^{2}}
 \left(\md r^2+r^2\md\Omega^2\right)
\]
with the scale of the body determined by $a(t)$, and matching it to a
generalized Vaidya spherically symmetric outside region
\[
 \md s^2=-\left(1-\frac{2 M(v,\chi)}{\chi}\right)\md v^2 
 +2\md v\md\chi+\chi^2\md\Omega^2
\]
with a function $M(v,\chi)$ determining the outside matter flux,
leads to the conditions
\begin{eqnarray}
 \chi(v) &=& Ra(t)/(1+R^2/4)\\
  2M &=& aR^3(\dot{a}^2+1)/(1+R^2/4)^3
\end{eqnarray}
at the matching surface. Here, $R$ is the coordinate value for $r$
where the interior solution is cut off.

The homogeneous interior can then be described by effective equations
including repulsive quantum correction terms as discussed in
Sec.~\ref{lqc} \cite{Collapse}. With such a term, the interior bounces
which, through the matching, also influences the exterior geometry and
its horizons. In the absence of effective equations for inhomogeneous
situations, the full outside behavior cannot be determined. But at
least in the neighborhood of the matching surface one can study the
formation and possible disappearance of horizons. A marginally trapped
spherical surface forms when $2M(v,\chi)=\chi$ is satisfied which,
with the matching conditions, implies $|\dot{a}|=(1-R^2/4)/R$ for the
interior.

Classically, $\dot{a}$ is unbounded and monotonic such that there is
always exactly one solution to the condition of a marginally trapped
surface at the matching surface. There is thus only a single horizon
covering the classical singularity.

\begin{figure}
\begin{center}
\includegraphics[height=5cm]{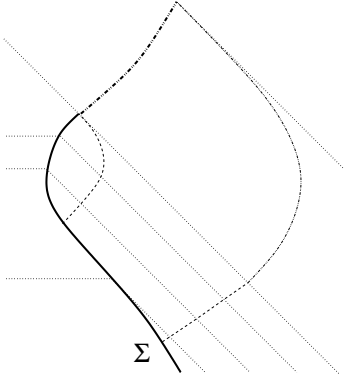}\hspace{1cm}
\parbox[b]{9cm}{\caption{Matching surface $\Sigma$ of a bouncing interior (left) matched to
an inhomogeneous outside region with evaporating horizons.
\label{Horizon}}}
\end{center}
\end{figure}

With repulsive quantum effects, the situation changes: First,
$\dot{a}$ starts to decrease before the bounce in the effective
dynamics, implying the existence of a second solution corresponding to
an inner horizon as illustrated in Fig.~\ref{Horizon}.  Secondly,
$\dot{a}$ is bounded and there are cases, depending on parameters and
initial conditions, without any solution for $2M(v,\chi)=\chi$. The
classical singularity is then replaced by a bounce which is covered by
horizons only for larger mass. This scenario thus indicates the
presence of a threshold for black hole formation.

\subsubsection{Midisuperspaces}

The quantum behavior across horizons can only be seen in inhomogeneous
models. For spherically symmetric ones, the loop representation leads
to states of the form \cite{SphSymm}
\begin{equation}
 |\psi\rangle= \sum_{\vec{k},\vec{\mu}}\psi(\vec{k}, \vec{\mu})
\mbox{{\unitlength=0.2mm\begin{picture}(200,10)(0,2)
    \put(0,5){\line(1,0){200}} \put(50,5){\circle*{5}}
    \put(100,5){\circle*{5}} \put(150,5){\circle*{5}}
 \put(50,-5){\makebox(0,0){$\mu_-$}}
 \put(100,-5){\makebox(0,0){$\mu$}}
 \put(150,-5){\makebox(0,0){$\mu_+$}}
 \put(25,10){\makebox(0,0){$\cdots$}}
 \put(75,12){\makebox(0,0){$\scriptstyle k_-$}}
 \put(125,12){\makebox(0,0){$\scriptstyle k_+$}}
 \put(175,10){\makebox(0,0){$\cdots$}}
\end{picture}} with $k_e\in{\mathbb Z}$, $0\leq\mu_v\in{\mathbb R}$}
\end{equation}
subject to coupled difference equations (one for each edge)
\begin{eqnarray}
 && \hat{C}_0(\vec{k})\psi(\ldots,k_-,k_+,\ldots)+
 \hat{C}_{R+}(\vec{k})\psi(\ldots,k_-,k_+-2,\ldots)\nonumber\\
 &&+\hat{C}_{R-}(\vec{k})\psi(\ldots,k_-,k_++2,\ldots)
 +\hat{C}_{L+}(\vec{k})\psi(\ldots,k_--2,k_+,\ldots)\nonumber\\
 &&+\hat{C}_{L-}(\vec{k})\psi(\ldots,k_-+2,k_+,\ldots)+\cdots =0
\end{eqnarray}
with coefficients which have been computed
explicitly \cite{SphSymmHam}. Also here, superspace is extended by the
freedom of (local) orientation: ${\rm sgn}\det E$ is determined by
${\rm sgn} k_e$.

Again, the quantum equations are {\em non-singular} \cite{SphSymmSing},
however much more crucially depending on the form (in particular on
possible zeros) of the coefficients $\hat{C}_{R\pm}(\vec{k})$.  Unlike
in homogeneous models, a {\em symmetric ordering} is required to
extend solutions. In isotropic models this removes any possible
dynamical initial conditions, but in less symmetric models the
solution space is still restricted. How strong the restrictions will
be has to be seen from a more detailed analysis of the resulting
initial/boundary value problem for difference equations.

Also unlike homogeneous models, the anomaly issue plays a bigger role:
coupled difference equations must be consistent with each other for a
well-defined initial/boundary value problem. While the anomaly issue
in this model is open as of now, the existence and uniqueness of
solutions in terms of suitable initial and boundary values has been
shown. This has to be revisited, however, for the issue of
semiclassical properties.  In this model, there are further
qualitatively different possibilities for the constraint
operator. While the above discussion was based on a fixed number of
labels (an operator not creating new spin network vertices), also
other variants exist. For such a choice, the number of degrees of
freedom would not be preserved when acting with the constraint, and a
different type of recurrence problem arises.  This freedom is
analogous to quantization choices in the full theory which makes it
possible to compare ambiguities and restrict choices by tight
mathematical consistency conditions and the physical viability of
scenarios.  So far, the detailed quantization is far from unique
(except for kinematical aspects), but there are characteristic and
robust generic effects.

\subsubsection{Quantum Horizons}

In midisuperspace models, it is also easier than in a full setting to
impose horizon conditions at the quantum level and study quantum
horizon dynamics. Isolated horizons \cite{HorRev} are particularly
useful because in spherical symmetry they simplify considerably in
Ashtekar variables \cite{Horizon}. This is useful for loop quantum
gravity where Ashtekar variables are basic, and here even the quantum
dynamics simplifies in the neighborhood of horizons. Moreover, the
simplifications occur approximately also for slowly evolving
horizons \cite{SlowHor} such that even dynamical situations are
accessible. Alternatively, direct quantizations of classical expansion
parameters have been formulated in
\cite{HWHorizon,BHCohHor,TriangBH} for fully dynamical
situations.

Conclusions drawn so far confirm quantum fluctuations of
horizons \cite{Horizon} as suggested often before by heuristic
arguments. One can also easily count the degrees of freedom of exactly
spherically symmetric horizons, but the symmetry reduction removes far
too many of them for a faithful counting of black hole entropy. This
is the one issue where symmetric models are clearly not reliable and
one has to use the full theory. Fortunately, even the full dynamics
simplifies if an isolated horizon is introduced as a boundary allowing
the correct counting of entropy for all astrophysically relevant black
holes \cite{ABCK:LoopEntro,IHEntro,RotatingBH}.

\section{Non-Uniqueness}

The framework of loop quantum gravity provides promising indications,
but so far cannot be seen as a complete theory due to a large amount
of ambiguities as they often occur in non-linear quantum theories
(discussed, e.g., in
\cite{Gaul,Ambig,ICGC,NPZRev,AlexAmbig}). This fact belongs
to a much broader issue about uniqueness in quantum gravity where
often two very different kinds are envisaged: a unique theory vs.\ a
unique solution (see also \cite{LivRev}). These are indeed very
different concepts as the uniqueness of a theory as such is not
testable even in principle and thus of metaphysical quality. When a
theory has a unique solution, however, its properties can be compared
with observations at least in principle.

In fact, both concepts may be contradictory for all we know so far:
there are ambiguities in loop quantum gravity and thus no unique
theory (although solutions may be restricted by dynamical initial
conditions giving some degree of uniqueness at the level of
solutions), while the supposedly unique string theory has a whole
landscape of potentially admissible solutions \cite{KKLT,StringStat}.
From a philosophical point of view, this situation has a precedent
which led to the following statement \cite{FN}:
\begin{quote}
 ``{\em ... a vast new panorama opens up for him, a
    possibility makes him giddy, mistrust, suspicion and fear of every
    kind spring up, belief in m[...], every kind of m[...], wavers,
    --- finally, a new demand becomes articulate. So let us give voice
    to this {\em new demand}: we need a {\em critique} of m[...]
    values, {\em the value of these values should itself, for once, be
      examined} ---}''
\end{quote}
Today, one may be tempted to complete the m-word by ``M-theory,'' but
in those days it was actually ``morality.'' Historically, philosophers
attempted to construct something which in the current terminology
could be called ``Grand Unified Morality'' or GUM, most widely known
in the form of Kant's categorical imperative. This was meant as a
unique theory, but did have too many solutions. Nietzsche's lesson was
that the overly idealistic approach had to be replaced by a rather
phenomenological one, where he studied the behavior (i.e.\ the
phenomenology of morality) in different cultures. Similarly, quantum
gravity may currently be at such a crossroads where idealism has to be
replaced by phenomenology.

\section{Conclusions}

Quantum effects are significant at small scales and lead to
qualitatively new behavior. Intuitively, this can be interpreted
as repulsive contributions to the gravitational force for which there
are several examples in the framework of loop quantum
gravity \cite{LivRev}. When such effects are derived rather than being
chosen with phenomenological applications in mind, it is by no means
guaranteed that their modifications take effect in the correct
regimes. This gives rise to many consistency checks such as those
discussed here in anisotropic and inhomogeneous models in the context
of black holes.  These effects, while not fixed in detail, are robust
and rather direct consequences of the loop representation, with
non-perturbativity and background independence being essential.

With these effects, the theory can resolve a variety of conceptual and
technical problems from basic effects, without the need to introduce
new ingredients. At the current stage we have a consistent picture of
the universe, including the classically puzzling situations of the big
bang and black holes, which is well-defined everywhere. From here, one
can use internal consistency and potential contact with observations
to constrain the remaining freedom and test the whole framework.


\end{document}